\documentclass[10pt]{article}

\usepackage{graphicx}%
\usepackage{amsmath,amssymb,amsthm,upref,bm}%
\usepackage{labelfig}%

\usepackage{xcolor}

\begin{document}

\baselineskip=4.4mm

\makeatletter

\newcommand{\E}{\mathrm{e}\kern0.2pt} 
\newcommand{\D}{\mathrm{d}\kern0.2pt}
\newcommand{\RR}{\mathbb{R}}
\newcommand{\CC}{\mathbb{C}}%
\newcommand{\ii}{\kern0.05em\mathrm{i}\kern0.05em}

\renewcommand{\Re}{\mathrm{Re}} 
\renewcommand{\Im}{\mathrm{Im}}

\def\bottomfraction{0.9}

\title{\bf On sloshing in containers with porous baffles}

\author{Nikolay Kuznetsov and Oleg Motygin}

\date{}

\maketitle

\vspace{-8mm}

\begin{center}
Laboratory for Mathematical Modelling of Wave Phenomena, \\ Institute for Problems
in Mechanical Engineering, Russian Academy of Sciences, \\ V.O., Bol'shoy pr. 61, St
Petersburg 199178, Russian Federation \\ E-mail: nikolay.g.kuznetsov@gmail.com,
o.v.motygin@gmail.com
\end{center}

\begin{abstract}
\noindent Sloshing eigenvalues are studied for containers with porous baffles
extending throughout the constant (possibly infinite) depth. The fluid transmission
across the baffles is described by Darcy's law, and so the spectral problem is
nonself-adjoint. In particular, much attention is paid to vertical cylinders with
circular walls and radial baffles. Explicit solutions are obtained, whose crucial
feature is that the corresponding pressure difference vanishes across the baffle.
These real solutions coincide with those that describe sloshing in the presence of
rigid baffles, thus demonstrating that at certain frequencies the damping efficiency
of porous baffles is the same as that of the rigid ones having the same
configuration.
\end{abstract}

\setcounter{equation}{0}

\section{Introduction}

Studies of fluid flows in regions bounded by porous surfaces date back to 1956,
when the groundbreaking article \cite{T} was published by G. I. Taylor in connection
with processes in a paper-making machine. However, his approach to this topic
remained unnoticed for almost 30 years; at last, Chwang \cite{C} applied it in his
theory of a porous wavemaker. In the paper \cite{LC}, this approach was extended to
the problems of scattering and radiation of water waves by porous barriers. However,
analytical solutions for scattering problems involving different vertical porous
barrier configurations in deep water were found only recently; see \cite{MS1}
and~\cite{MS2}. In the case of finite depth, the only obtained analytical solution
describes the wave interaction with a single vertical, porous barrier extending
throughout a two-dimensional layer~\cite{MJ}. During the past few years, sloshing in
containers with porous baffles attracted much attention; see the papers \cite{NS},
\cite{PC}, \cite{CK}, \cite{Tu} and references cited therein.

In this paper, our aim is to construct examples of explicit solutions to the
nonself-adjoint sloshing problem in containers with porous baffles. The fluid
transmission across such a baffle is described by the widely used Darcy's law.
According to this model (see \cite{Tu} and references cited therein), the fluid's
velocity at the baffle is directly proportional to the pressure difference across
the baffle itself. A rectangular container with the baffle dividing it into two
symmetric parts is considered first. Then the results obtained in \cite{KM} (they
concern sloshing of an inviscid, incompressible, heavy fluid in vertical cylinders
with circular walls and rigid, radial baffles) are accommodated to porous baffles.
The obtained solutions are real valued (an unusual feature that is specific for this
nonself-adjoint problem), which demonstrates that the damping efficiency of porous
baffles at some frequencies is the same as that of rigid ones having the same
configuration.

The paper's plan is as follows. Sloshing in rectangular containers is considered in
sect.~2. Sloshing in a vertical circular cylinder with a radial baffle extending
from the cylinder's axis to the wall is considered in sect.~3. The similar results
for an annular container with a radial baffle connecting walls are presented in
sect.~4. More examples of explicit solutions in a circular cylinder are given in
sect.~5; they concern multiple radial baffles in a circular cylinder. Concluding
remarks and discussion are given in sect.~6.

\section{Rectangular containers}

It is convenient to consider the following rectangular container with the midplane
baffle in the $(y,z)$-plane:
\[ W_1 = \bigl\{(x,y,z): \, x \in (-a, 0) \cup (0, a), \, y \in (0, b), \, z \in (
-h,0)\bigr\} .
\]
Here $a, b > 0$, whereas $h \in (0, \infty]$ to admit the case of infinite depth.
The fluid transmission across the porous baffle $L = \{x=0, \, y \in (0, b), \, z
\in (-h,0)\}$ is described by Darcy's law. Its general mathematical formulation is
given in \cite[p.~22-5]{Tu}; moreover, see \cite[p.~22-7]{Tu} for its specific form
concerning normal modes of free oscillations. Under the usual hydrodynamic
assumptions listed below, sloshing normal modes and ``eigenfrequencies'' in the case
of finite $h$ are sought using eigenfunctions and eigenvalues, respectively, of the
following problem:
\begin{align}
& \phi_{xx} +  \phi_{yy} + \phi_{zz} = 0 \ \ \mbox{in} \ W_1 , \label{slosh1r1} \\ &
\phi_z = \omega^2 \phi / g \ \ \mbox{on} \ F = \{x \in (-a, 0) \cup (0, a) , \, y \in
(0, b) , \, z = 0\}, \label{slosh2r1} \\ & \phi_z = 0 \ \ \mbox{on} \ B = \{x \in
(-a, 0) \cup (0, a) , \, y \in (0, b) , \, z = -h\}, \label{slosh3r1} \\ & \phi_x =
0 \ \ \mbox{on} \ S_x = \{x = \pm a, \, y \in (0, b), \, z \in (-h, 0)\},
\label{slosh4r1x} \\ & \phi_y = 0 \ \ \mbox{on} \ S_y = \{x \in (-a, 0) \cup (0, a) ; 
\, y = 0, b; \, z \in (-h, 0)\}, \label{slosh4r1y} \\ & \phi_x (-
0, y, z) = \phi_x (+0, y, z) = \ii \beta \, \omega [ \phi (+0, y, z) - \phi (-0, y,
z) ] \ \ \mbox{on} \ L . \label{slosh5r1}
\end{align}
In this formulation, $g > 0$ stands for the constant acceleration due to gravity;
the velocity potential $\phi$ is assumed to be complex valued and belonging to the
Sobolev space $H^1 (W_1)$; the latter means that the energy of oscillations is
finite. Also, $\omega$ is assumed to be complex as well as the coefficient $\beta =
\beta_r + \ii \beta_i \ (\beta_r, \beta_i > 0)$, which characterizes the porosity of
the baffle; its real part represents the resistance effect of the baffle against the
flow, whereas the imaginary part represents the inertial effect of the fluid in the
baffle; see \cite{Tu} and references cited therein. Thus, the spectral problem
\eqref{slosh1r1}--\eqref{slosh5r1} is nonself-adjoint due to the last relation in
\eqref{slosh5r1}. In the case when $L$ is rigid (of course, it must not extend
throughout the depth, thus dividing the fluid into two parts), one has to put zero
in place of the last expression, which makes the problem self-adjoint with the real
spectral parameter $\nu = \omega^2 / g$, where $\omega$ is the radian sloshing
frequency; see, for example, \cite{EM}.

As usual, \eqref{slosh1r1} is a consequence of the continuity equation for the
irrotational motion; \eqref{slosh2r1} follows from Bernoulli's equation by
linearization on the horizontal mean free surface $F$. Relations \eqref{slosh3r1},
\eqref{slosh4r1x} and \eqref{slosh4r1y} are the no flow conditions on the rigid
bottom $B$ and the vertical walls $S_x$ and $S_y$ respectively. The fluid's normal
velocity is the same on both sides of $L$ according to the first relation
\eqref{slosh5r1}, whereas the second of these relations is Darcy's law for the
porous baffle $L$. 

As in the case of a rectangular container without baffle, separation of variables
yields the following explicit solutions of problem
\eqref{slosh1r1}--\eqref{slosh5r1}:
\begin{equation}
\phi_{n,m} (x, y, z) = A \cos (n \pi x / a) \cos (m \pi x / b) \cosh \kappa_{n,m} (z
+ h) . \label{ur1}
\end{equation}
Here $A$ is an arbitrary non-zero constant; $n = 0,1,2,\dots$, $m = 0,1,2,\dots$,
and they are not equal to zero simultaneously (this excludes the non-physical
constant solution), whereas
\begin{equation}
\kappa_{n,m} =  \bigl[ (n \pi / a)^2 + (m \pi / b)^2 \bigr]^{1/2}. \label{kappa}
\end{equation}
It is straightforward to verify that $\phi_{n,m}$ satisfies relations
\eqref{slosh1r1}--\eqref{slosh4r1y} provided 
\[ \omega^2_{n,m} / g = \kappa_{n,m} \tanh (\kappa_{n,m} \, h) ,
\]
where $\kappa_{n,m}$ is given by formula \eqref{kappa}. Moreover, since the
rightmost expression in \eqref{slosh5r1} vanishes, the symmetry of $W_1$ about the
$(y,z)$-plane guarantees that all conditions \eqref{slosh5r1} are fulfilled.

The obtained solution demonstrates that at every frequency $\sqrt{g \,
\kappa_{n,m}}$ sloshing in $W_1$ with any porous baffle $L$ is equivalent to
synchronous, symmetric sloshing in both halves of $W_1$ separated by the rigid
baffle.

In the case of infinite depth when decaying of $\phi$ is required as $z \to -\infty$
instead of condition \eqref{slosh3r1}, the solutions of problem
\eqref{slosh1r1}--\eqref{slosh5r1} analogous to \eqref{ur1} are as follows:
\[ \phi_{n,m} (x, y, z) = A \cos (n \pi x / a) \cos (m \pi x / b) \, \E^{\kappa_{n,m} z}
\ \ \mbox{and} \ \omega^2_{n,m} = g \, \kappa_{n,m} .
\]

In the same way, one obtains normal modes and sloshing frequencies in the case when
a rectangular container has two porous baffles located at orthogonal vertical
midplanes.

\begin{figure}[t]
\centering\vspace{1.5mm}
 \SetLabels
 \L (0.67*0.797) $x$\\
 \L (0.4*0.835) $y$\\
 \L (0.55*0.97) $z$\\
 \L (0.31*0.74) $F$\\
 \L (0.96*0.45) $S$\\
 \L (0.51*0.04) $B$\\
 \L (0.69*0.44) $L$\\
 \L (0.31*0.125) $r$\\
 \L (0.46*0.1625) $\theta$\\
 \L (-0.03*0.45) $h$\\
 \endSetLabels
 \leavevmode\AffixLabels{\includegraphics[width=40mm]{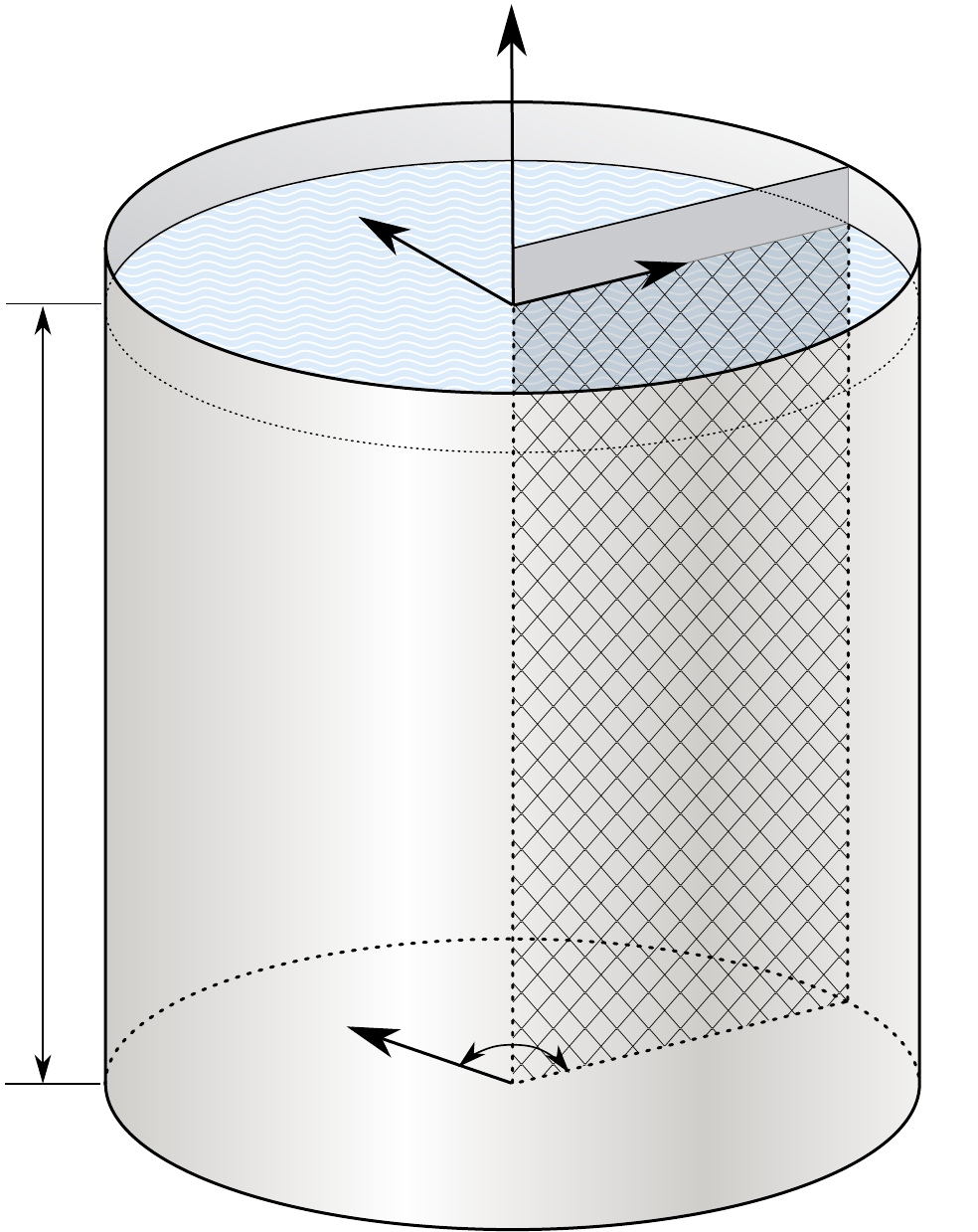}}
 \caption{A sketch of the circular cylinder $W$ with a radial baffle.}
 \label{fig:def1}
\end{figure}
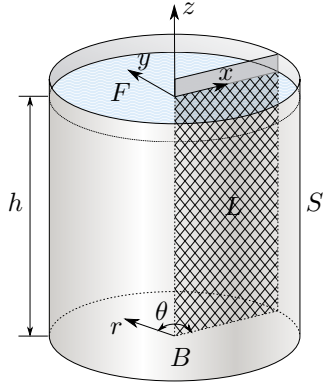

\section{Vertical circular cylinder with a radial baffle}

It is convenient to take the vertical circular cylinder without a baffle in the
form:
\[ W_0 = \bigl\{(x,y,z): \, {x^2 + y^2 < 1}, \, z \in (-h,0)\bigr\} , \]
where $h \in (0, \infty]$. Using separation of variables \cite[formulae (6), (7)]{KM}, the
sloshing eigenvalues of the spectral parameter $\nu = \omega^2 / g$ ($\omega$ is the
radian sloshing frequency) are obtained explicitly for this domain; see \cite{KM},
formula (11) and fig.~2. Moreover, these eigenvalues are compared in \cite{KM} with
the explicit eigenvalues describing sloshing in the domain
\[ W = \{(r, \theta, z): \, r \in (0, 1) , \, \theta \in (0, 2 \pi) , \, z \in
(-h,0)\} ,
\]
in the case when the rectangular baffle $L = \{(r, 0, z): \, r \in [0, 1] , \, z \in
[-h,0]\}$ immersed into $W_0$ (see fig.~1) is assumed to be rigid. Here and below $r$
is the first component of the cylindrical coordinates $(r, \theta, z)$ such that $x
= r \cos \theta$ and $y = r \sin \theta$. The damping effect due to the rigid baffle
$L$ is shown in \cite[fig.~4]{KM}.

In the present paper, we consider sloshing in $W$ in the case when the baffle $L$ is
porous, and so sloshing modes and frequencies in the case of finite $h$ are sought
using eigenfunctions and eigenvalues, respectively, of the following problem:
\begin{align}
& \phi_{xx} +  \phi_{yy} + \phi_{zz} = 0 \ \ \mbox{in} \ W , \label{slosh1} \\ &
\phi_z = \omega^2 \phi / g \ \ \mbox{on} \ F = \{(r, \theta, 0): \, r \in (0, 1) , \,
\theta \in (0, 2 \pi)\}, \label{slosh2} \\ & \phi_z = 0 \ \ \mbox{on} \ B = \{(r,
\theta, -h): \, r \in (0, 1) , \, \theta \in (0, 2 \pi)\}, \label{slosh3} \\ &
\phi_r = 0 \ \ \mbox{on} \ S = \{(1, \theta, z): \, \theta \in (0, 2 \pi), \, z \in
(-h, 0)\},
\label{slosh4} \\ & \phi_y (r, 0, z) = \phi_y (r, 2 \pi, z) = \ii \beta \, \omega [
\phi (r, 0, z) - \phi (r, 2 \pi, z) ] \ \ \mbox{for} \ r \in (0, 1), \ z \in (-h, 0)
. \label{slosh5}
\end{align}
As in sect. 2, the velocity potential $\phi$ is assumed to be complex valued and
belonging to the Sobolev space $H^1 (W)$; also, the spectral parameter $\omega$ is
assumed to be complex as well as the constant coefficient $\beta = \beta_r + \ii
\beta_i \ (\beta_r, \beta_i > 0)$, which characterizes the porosity of the baffle.

In the case of rigid $L$, the sloshing problem was considered in
\cite[sect.~III\,B]{KM}; it has the same relations \eqref{slosh1}--\eqref{slosh4},
but the last condition is just
\begin{equation}
\phi_y (r, 0, z) = \phi_y (r, 2 \pi, z) = 0 \ \ \mbox{for} \ r \in (0, 1), \ z \in
(-h, 0) . \label{slosh5'}
\end{equation}
This condition follows from \eqref{slosh5} either by setting $\beta = 0$ or
requiring
\begin{equation}
\phi_{n,s} (r, 0, z) - \phi_{n,s} (r, 2 \pi, z) = 0 . \label{req}
\end{equation}
Therefore, for obtaining explicit solutions of problem
\eqref{slosh1}--\eqref{slosh5} with an arbitrary complex $\beta$ we notice that
these coincide with certain solutions obtained in \cite[sect.~III\,B]{KM} under the
assumption that $L$ is rigid; namely, they have the form:
\begin{equation}
\phi_{n,s} (r, \theta, z) = u_{n,s} (r, \theta) \cosh k_{n,s} (z + h) \ \ \mbox{and}
\ \ \omega^2_{n,s} = g \, k_{n,s} \tanh k_{n,s} h , \label{u}
\end{equation}
where $n = 0,1,2,\dots$ and $s = 1,2,\dots$. Furthermore,
\begin{equation}
u_{n,s} (r, \theta) = A J_n \bigl( k_{n,s} \, r \bigr) \cos n \theta \label{com}
\end{equation}
with an arbitrary non-zero constant $A$; here $J_n$ is the Bessel function of order
$n$ and $k_{n,s}$ is the $s$th positive zero of its derivative $J'_n$.

\begin{figure}[t]
\centering\vspace{1.25mm}
 \SetLabels
 \L (0.97*0.01) $n$\\
 \L (0.132*0.205) $\tilde \nu_1$\\
 \L (0.217*0.29) $\tilde \nu_2 = \nu_1$\\
 \L (0.303*0.3675) $\tilde \nu_3$\\
 \L (0.39*0.445) $\tilde \nu_4 = \nu_2$\\
 \L (0.476*0.52) $\tilde \nu_5$\\
 \L (0.05*0.547) $\tilde \nu_6 = \nu_3$\\
 \L (0.565*0.595) $\tilde \nu_7 = \nu_4$\\
 \L (0.133*0.645) $\tilde \nu_8$\\
 \L (0.65*0.667) $\tilde \nu_9$\\
 \L (0.735*0.74) $\tilde \nu_{10} = \nu_5$\\
 \L (0.22*0.742) $\tilde \nu_{11} = \nu_6$\\
 \L (0.82*0.81) $\tilde \nu_{12}$\\
 \L (0.304*0.83) $\tilde \nu_{13}$\\
 \L (0.908*0.882) $\tilde \nu_{14} = \nu_7$\\
 \L (0.393*0.918) $\tilde \nu_{15} = \nu_8$\\
 \L (0.995*0.952) $\tilde \nu_{16}$\\
 \L (0.05*0.96) $\tilde \nu_{17} = \nu_9$\\
 \endSetLabels
 \leavevmode\AffixLabels{ \includegraphics[width=100mm]{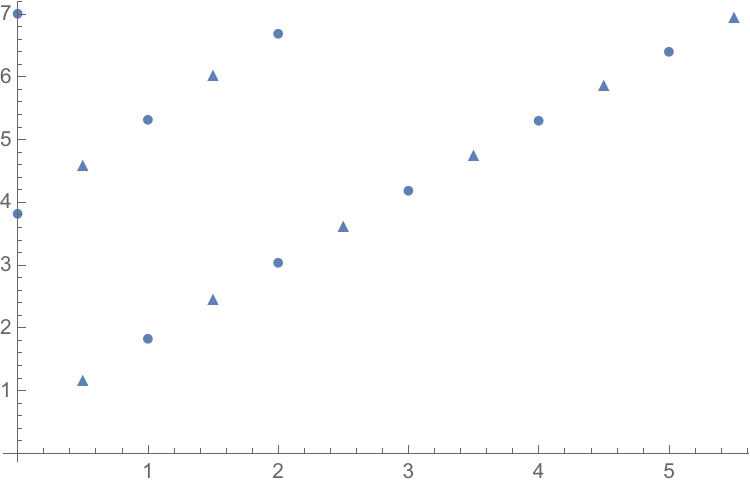}}
 \vspace{-2mm}
 \caption{Sloshing eigenvalues for circular, infinitely deep cylinders with radial
 baffles: $\tilde \nu_m$ for the rigid baffle, $m=1,2,\ldots,17$; $\nu_k$ for the
 porous baffle, $k=1,2,\ldots,9$.}
 \label{fig:circ_with_baffle}
\end{figure}
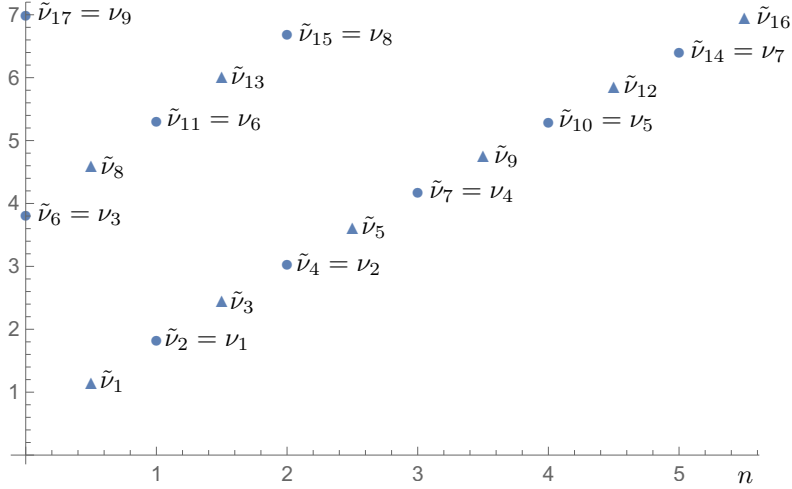

It is straightforward to verify that $\phi_{n,s}$ satisfies relations
\eqref{slosh1}--\eqref{slosh4} provided $\nu_{n,s} = \omega^2_{n,s} / g$ (it is
convenient to use the same parameter as in the case of rigid baffle) is given by the
second formula \eqref{u}. Moreover, the dependence on $\cos n \theta$ with integer
$n$ in \eqref{com} is crucial for conditions \eqref{slosh5} to be valid; indeed, it
implies that \eqref{req} is fulfilled, and so \eqref{slosh5} reduces to the rigid
baffle condition \eqref{slosh5'}. The latter also holds in view of the dependence on
$\cos n \theta$ in \eqref{com}.

In the case of infinite depth when decaying of $\phi$ is required as $z \to -\infty$
instead of condition \eqref{slosh3}, the analogous solutions of problem
\eqref{slosh1}--\eqref{slosh5} are as follows:
\begin{equation}
\phi_{n,s} (r, \theta, z) = u_{n,s} (r, \theta) \, \E^{ k_{n,s} z } \ \ \mbox{and} \
\ \nu_{n,s} = k_{n,s} , \ \ n = 0,1,2,\dots , \ s = 1,2,\dots . \label{u'}
\end{equation}
They are slightly different from \eqref{u}, but $u_{n,s}$ is the same as above; see
\eqref{com}.

It is worth noting that formulae \eqref{u}, \eqref{u'} and \eqref{com} also give
solutions of problem \eqref{slosh1}--\eqref{slosh4} and \eqref{slosh5'} for all
positive, half-integer values of $n$; see \cite[sect.~III\,B]{KM}. However,
\[ \phi_{n,s} (r, 0, z) - \phi_{n,s} (r, 2 \pi, z) \neq 0 
\]
in this case, which means that $\phi_{n,s}$ does not solve problem
\eqref{slosh1}--\eqref{slosh5} when $n$ is half-integer. Indeed, the last relation
violates the second condition \eqref{slosh5} because $\phi_{n,s}$ still satisfies
\eqref{slosh5'} when $n$ is half-integer. This fact is illustrated in fig.~2, where
$\tilde \nu_m$, $m=1,2,\ldots,17$ (marked by both triangles for half-integer $n$ and
bullets for integer $n$) are simple sloshing eigenvalues for the circular,
infinitely deep cylinder when the baffle is rigid, whereas $\nu_k$, $k=1,2,\ldots,9$
(marked only by bullets) correspond to the porous baffle.

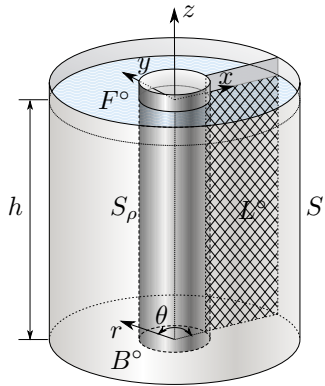
\begin{figure}[b]
\centering 
 \SetLabels
 \L (0.67*0.797) $x$\\
 \L (0.4*0.8375) $y$\\
 \L (0.55*0.975) $z$\\
 \L (0.27*0.73) $F^\circ$\\
 \L (0.96*0.45) $S$\\
 \L (0.31*0.45) $S_\rho$\\
 \L (0.31*0.042) $B^\circ$\\
 \L (0.735*0.44) $L^\circ$\\
 \L (0.31*0.125) $r$\\
 \L (0.46*0.1625) $\theta$\\
 \L (-0.03*0.45) $h$\\
 \endSetLabels
 \leavevmode\AffixLabels{\includegraphics[width=40mm]{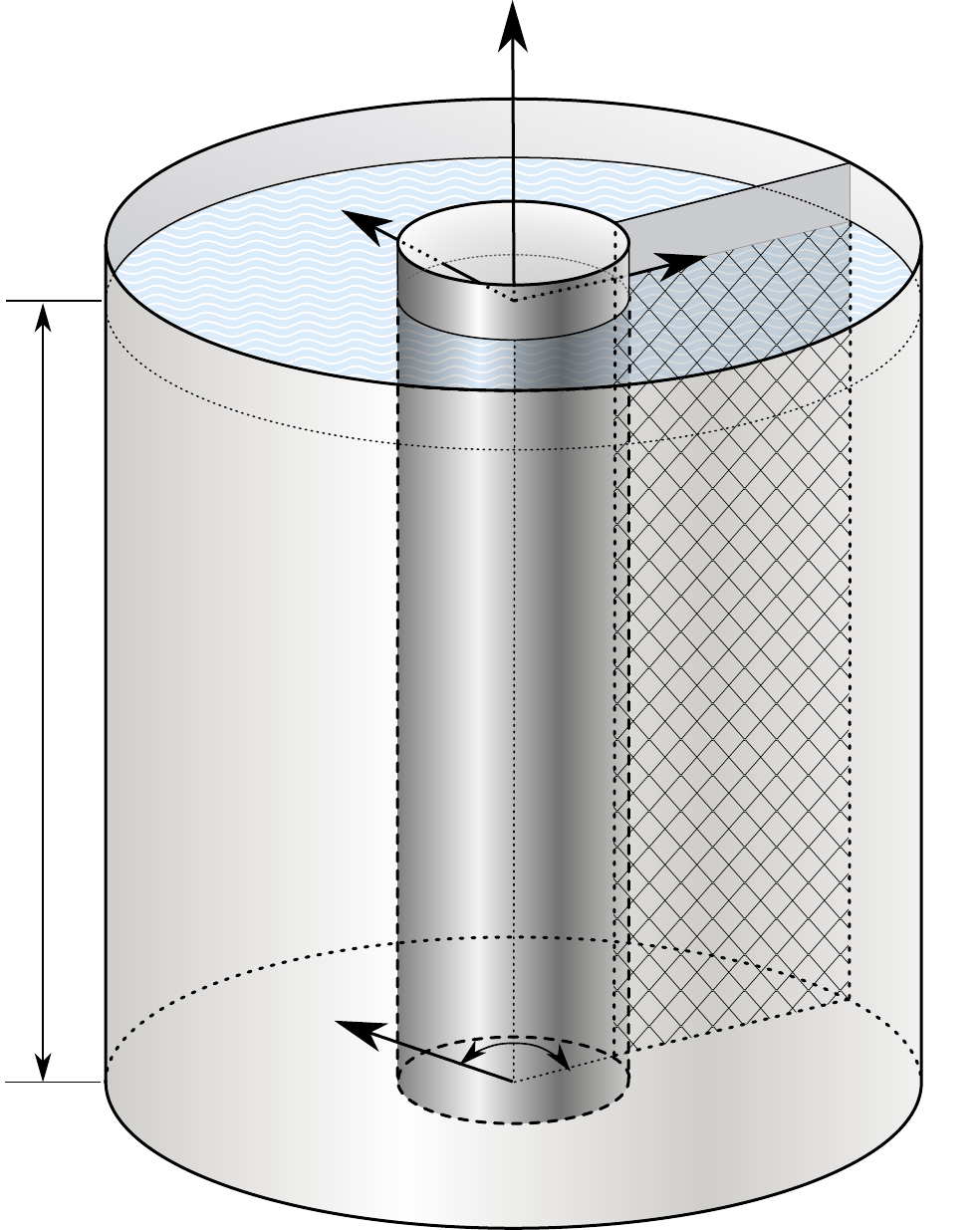}} 
 \caption{A sketch of the annular cylinder $W^\circ$ with a radial baffle.}
 \label{fig:def2}
\end{figure}

\section{Annular cylinder with a radial baffle}

It is convenient to take the vertical annular cylinder without a baffle in the form:
\[ W_0^\circ = \bigl\{(x,y,z): \, \rho < x^2 + y^2 < 1, \, z \in (-h,0)\bigr\} , \]
where $\rho \in (0, 1)$ and $h \in (0, \infty]$. Again, the sloshing eigenvalues of
the spectral parameter $\nu = \omega^2 / g$ are obtained explicitly for this domain
(see \cite[formulae (21) and (22)]{KM}) and compared with the explicit eigenvalues
describing sloshing in the domain
\[ W^\circ = \{(r, \theta, z): \, r \in (\rho, 1) , \, \theta \in (0, 2 \pi) , \, z
\in (-h,0)\} ,
\]
which contains the rectangular rigid baffle $L^\circ = \{(r, 0, z): \, r \in [\rho,
1] , \, z \in [-h,0]\}$; see fig.~3. The presence of the baffle $L^\circ$ diminishes
sloshing eigenvalues comparing with those in the same container without baffle; the
difference is substantial for the lowest eigenvalue; see \cite[figs.~8 and 9]{KM}.

Let us consider sloshing in $W^\circ$ in the case when the baffle $L^\circ$ is
porous, that is, the fluid transmission across $L^\circ$ is described by Darcy's
law. Similar to \eqref{slosh1}--\eqref{slosh5}, we have the following problem for
the complex valued velocity potential $\phi$:
\begin{align}
& \phi_{xx} +  \phi_{yy} + \phi_{zz} = 0 \ \ \mbox{in} \ W^\circ , \label{slosh1a}
\\ & \phi_z = \omega^2 \phi / g \ \ \mbox{on} \ F^\circ = \{(r, \theta, 0): \, r \in
(\rho, 1) , \, \theta \in (0, 2 \pi)\}, \label{slosh2a} \\ & \phi_z = 0 \ \
\mbox{on} \ B^\circ = \{(r, \theta, -h): \, r \in (\rho, 1) , \, \theta \in (0, 2
\pi)\}, \label{slosh3a} \\ & \phi_r = 0 \ \ \mbox{on} \ S \cup S_\rho, \ \ S_\rho =
\{(\rho, \theta, z): \, \theta \in (0, 2 \pi), \, z \in (-h, 0)\}, \label{slosh4a}
\\ & \phi_y (r, 0, z) = \phi_y (r, 2 \pi, z) = \ii \beta \, \omega [ \phi (r, 0, z)
- \phi (r, 2 \pi, z) ] \ \ \mbox{for} \ r \in (\rho, 1), \ z \in (-h, 0) .
\label{slosh5a}
\end{align}

Again, explicit solutions of problem \eqref{slosh1a}--\eqref{slosh5a} with an
arbitrary complex $\beta$ coincide with certain solutions obtained in
\cite[sect.~IV\,B]{KM} under the assumption that $L^\circ$ is rigid; that is, using
the condition
\begin{equation}
\phi_y (r, 0, z) = \phi_y (r, 2 \pi, z) = 0 \ \ \mbox{for} \ r \in (\rho, 1),
\ z \in (-h, 0) \label{slosh5'a}
\end{equation}
instead of \eqref{slosh5a}. These solutions have the same form as those in sect.~3:
\begin{equation}
\phi_{n,s}^\circ (r, \theta, z) = u_{n,s}^\circ (r, \theta) \cosh k_{n,s}^\circ (z +
h) \ \ \mbox{and} \ \ \big(\omega_{n,s}^\circ\big)^2 / g = \nu_{n,s}^\circ =
k_{n,s}^\circ \tanh k_{n,s}^\circ h , \label{ua}
\end{equation}
where $n = 0,1,2,\dots$ and $s = 1,2,\dots$. However, $k_{n,s}^\circ$ is the $s$th
positive root of the equation 
\[ J'_n (k_{n,s}^\circ \rho) \, Y'_n (k_{n,s}^\circ) - J'_n (k_{n,s}^\circ) \, Y'_n (k_{n,s}^\circ \rho) = 0 ,
\]
where $Y_n$ is the Bessel function of the second kind, and
\begin{equation}
u_{n,s}^\circ (r, \theta) = A R_{n,s}^\circ (k_{n,s}^\circ , r) \cos n \theta
\label{coma}
\end{equation}
instead of \eqref{com}. Here $A$ is an arbitrary non-zero constant, whereas
\[ R_{n,s}^\circ (k_{n,s}^\circ , r) = H_{n,s} (k_{n,s}^\circ , r) / H_{n,s}
(k_{n,s}^\circ , 1) 
\]
and
\[ H_{n,s} (k_{n,s}^\circ , r) = J_n (k_{n,s}^\circ r) \, Y'_n (k_{n,s}^\circ) -
J'_n (k_{n,s}^\circ) \, Y_n (k_{n,s}^\circ r) .
\]

It is straightforward to verify that $\phi_{n,s}^\circ$ satisfies relations
\eqref{slosh1a}--\eqref{slosh4a} provided $\nu_{n,s}^\circ$ is given by the second
formula \eqref{ua}. Moreover, the dependence on $\cos n \theta$ with integer $n$ in
\eqref{coma} is crucial for conditions \eqref{slosh5a} to be valid; indeed, it
implies that
\[ \phi_{n,s}^\circ (r, 0, z) - \phi_{n,s}^\circ (r, 2 \pi, z) = 0 ,
\]
and so \eqref{slosh5a} reduces to the rigid baffle condition \eqref{slosh5'a}. The
latter also holds in view of the dependence on $\cos n \theta$ in \eqref{coma}.

In the case of infinite depth when decaying of $\phi^\circ$ is required as $z \to
-\infty$ instead of condition \eqref{slosh3a}, the analogous solutions of problem
\eqref{slosh1a}--\eqref{slosh5a} are as follows:
\begin{equation}
\phi_{n,s}^\circ (r, \theta, z) = u_{n,s}^\circ (r, \theta) \, \E^{ k_{n,s}^\circ z
} \ \ \mbox{and} \ \ \nu_{n,s}^\circ = k_{n,s}^\circ , \ \ n = 0,1,2,\dots , \ s =
1,2,\dots . \label{u'a}
\end{equation}
They are slightly different from \eqref{ua}, but $u_{n,s}^\circ$ is the same as
above; see \eqref{coma}.

\begin{figure}[t]
\centering
 \SetLabels
 \L (0.115*0.13) $\tilde \nu_1^\circ$\\
 \L (0.196*0.214) $\tilde \nu_2^\circ=\nu_1^\circ$\\
 \L (0.275*0.295) $\tilde \nu_3^\circ$\\
 \L (0.356*0.38) $\tilde \nu_4^\circ=\nu_2^\circ$\\
 \L (0.436*0.456) $\tilde \nu_5^\circ$\\
 \L (0.517*0.535) $\tilde \nu_6^\circ=\nu_3^\circ$\\
 \L (0.595*0.61) $\tilde \nu_7^\circ$\\
 \L (0.675*0.685) $\tilde \nu_8^\circ=\nu_4^\circ$\\
 \L (0.754*0.758) $\tilde \nu_9^\circ$\\
 \L (0.834*0.827) $\tilde \nu_{10}^\circ=\nu_5^\circ$\\
 \L (0.035*0.83) $\tilde \nu_{11}^\circ$\\
 \L (0.035*0.78) $=\nu_6^\circ$\\
 \L (0.115*0.838) $\tilde \nu_{12}^\circ$\\
 \L (0.195*0.853) $\tilde \nu_{13}^\circ$\\
 \L (0.195*0.803) $=\nu_7^\circ$\\
 \L (0.275*0.88) $\tilde \nu_{14}^\circ$\\
 \L (0.913*0.898) $\tilde \nu_{15}^\circ$\\
 \L (0.355*0.917) $\tilde \nu_{16}^\circ$\\
 \L (0.355*0.867) $=\nu_8^\circ$\\
 \L (0.435*0.961) $\tilde \nu_{17}^\circ$\\
 \L (0.992*0.967) $\tilde \nu_{18}^\circ=\nu_9^\circ$\\
 \L (1*0.01) $n$\\
 \endSetLabels 
 \leavevmode\kern-2mm\AffixLabels{\includegraphics[width=100mm]{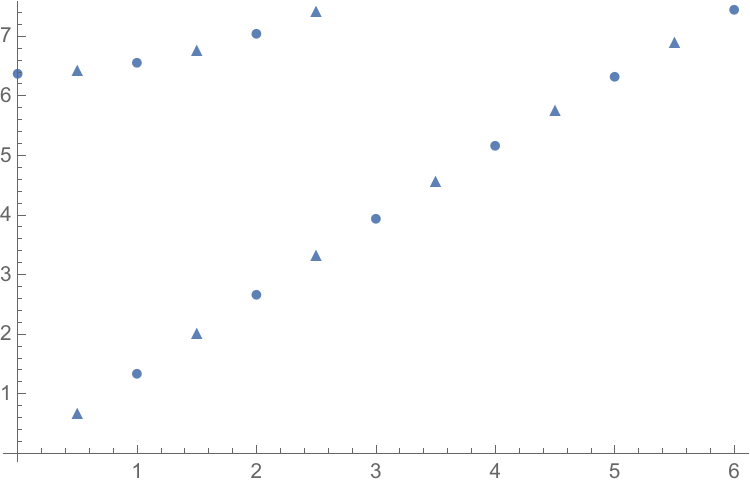}}
 \vspace{-2mm}
 \caption{Sloshing eigenvalues for annular ($\rho=1/2$), infinitely deep cylinders
 with radial baffles: $\tilde \nu_m^\circ$ for the rigid baffle, $m=1,2,\ldots,18$;
 $\nu_k^\circ$ for the porous baffle, $k=1,2,\ldots,9$.}
 \label{fig:ann}
\end{figure}

Again, formulae \eqref{ua}, \eqref{u'a} and \eqref{coma} also give solutions of
problem \eqref{slosh1a}--\eqref{slosh4a} and \eqref{slosh5'a} for all positive,
half-integer values of $n$; see \cite[sect.~IV\,B]{KM}. However,
\[ \phi_{n,s}^\circ (r, 0, z) - \phi_{n,s}^\circ (r, 2 \pi, z) \neq 0 
\]
in this case, which means that $\phi_{n,s}^\circ$ does not solve problem
\eqref{slosh1a}--\eqref{slosh5a} when $n$ is half-integer. Indeed, the last relation
violates the second condition \eqref{slosh5a} because $\phi_{n,s}^\circ$ still
satisfies \eqref{slosh5'a} when $n$ is half-integer. This fact is illustrated in
fig.~4, where $\tilde \nu_m^\circ$, $m=1,2,\ldots,18$ (marked by both triangles for
half-integer $n$ and bullets for integer $n$) are simple sloshing eigenvalues for
the annular, infinitely deep cylinder when the baffle is rigid, whereas
$\nu_k^\circ$, $k=1,2,\ldots,9$ (marked only by bullets) correspond to the porous
baffle.

\section{Radial baffles: more explicit solutions}

In this section, we begin with examples of explicit solutions that describe sloshing
in domains of the form:
\[ W_m = \biggl\{(r, \theta, z): \, r \in (0, 1) , \, \theta \in \biggl(0, \frac{2 \pi}{m}
\biggr) \cup \biggl(\frac{2 \pi}{m}, \frac{4 \pi}{m} \biggr) \cup \dots \cup \biggl(2
\pi \frac{m-1}{m}, 2 \pi \biggr) , \, z \in (-h,0) \biggr\} .
\]
They contain $m$ rectangular, radial baffles assumed to be porous and dividing $W_0$
into $m=2,3,\dots$ equal parts. Let $F_m$, $B_m$ and $S_m$ denote the free surface,
bottom and circular parts of the boundary $\partial W_m$, respectively. Then the
boundary value problem describing sloshing modes and frequencies in $W_m$ consists
of the same relations \eqref{slosh1}--\eqref{slosh4}, but fulfilled in $W_m$ and on
$F_m$, $B_m$, $S_m$, respectively; these must be complemented by the following set
of coupling conditions:
\begin{equation}
\frac{1}{r} \phi_\theta \biggl(r, \frac{2 \pi j}{m} - 0, z \biggr) = \frac{1}{r}
\phi_\theta \biggl(r, \frac{2 \pi j}{m} + 0, z \biggr) = \ii \beta \, \omega \biggl[
\phi \biggl(r, \frac{2 \pi j}{m} + 0, z \biggr) - \phi \biggl(r, \frac{2 \pi j}{m} - 0,
z \biggr) \biggr] , \label{slosh5s}
\end{equation}
which must hold for $r \in (0, 1)$, $z \in (-h, 0)$ and $j=1,\dots,m-1$. One more
coupling condition is as follows:
\begin{equation}
r^{-1} \phi_\theta (r, 0, z ) = r^{-1} \phi_\theta (r, 2 \pi, z ) = \ii \beta \,
\omega [ \phi (r, 0, z ) - \phi (r, 2 \pi, z ) ] . \label{slosh5s'}
\end{equation}

Similar to \eqref{u} and \eqref{com}, explicit solutions describing sloshing in
$W_m$ for any $m$ have the following form in the case of finite depth $h$:
\begin{equation}
\phi_{n,s}^{(m)} (r, \theta, z) = u_{n,s}^{(m)} (r, \theta) \cosh k_{m n,s} (z + h)
\ \ \mbox{and} \ \ \nu_{n,s}^{(m)} = k_{m n,s} \tanh k_{m n,s} h . \label{u4}
\end{equation}
Here $n = 0,1,2,\dots$, $s = 1,2,\dots$ and
\begin{equation}
u_{n,s}^{(m)} (r, \theta) = A J_{m n} \bigl( k_{m n,s} \, r \bigr) \cos (m n \theta) ,
\label{com4}
\end{equation}
where $A$ is an arbitrary non-zero constant and $k_{m n,s}$ is the $s$th positive
zero of $J'_{m n}$. Again formula \eqref{com4} guarantees that the pressure
difference vanishes across each baffle, and so \eqref{slosh5s} and \eqref{slosh5s'}
reduce to the rigid baffle condition which is fulfilled by \eqref{com4}.

An essential notice is that $\phi_{0,s}^{(m)} (r, z)$ with $u_{0,s}^{(m)} (r) = A
J_{0} \bigl( k_{0,s} \, r \bigr)$ serves as the explicit sloshing solution in any
$\hat W_m$ for any $s = 1,2,\dots$. This domain distinguishes from $W_m$ as follows;
it is $W_0$ divided into $m=2,3,\dots$ parts by an {\it arbitrary} distribution of
porous, radial baffles extending from the cylinder's axis to the circular wall.
Indeed, the absence of dependence on $\theta$ yields that the pressure difference
vanishes across any baffle.

Thus, we see that the described partitions of the domain $W_0$ into $m$ sectorial
parts by porous baffles has very strong effect on sloshing at every frequency
defined by $\nu_{n,s}^{(m)}$. Indeed, sloshing in each separate part is, in fact,
independent from oscillations in the rest of the whole cylindrical container. It is
interesting whether a similar effect happens in some container of other geometry.

\begin{figure}[t]
\centering\vspace{1.25mm}
 \SetLabels
 \L (-0.03*0.95) $\nu_{n,s}^{(m)}$\\
 \L (0.99*-0.01) $m n$\\
 \endSetLabels
 \leavevmode\AffixLabels{ \includegraphics[width=100mm]{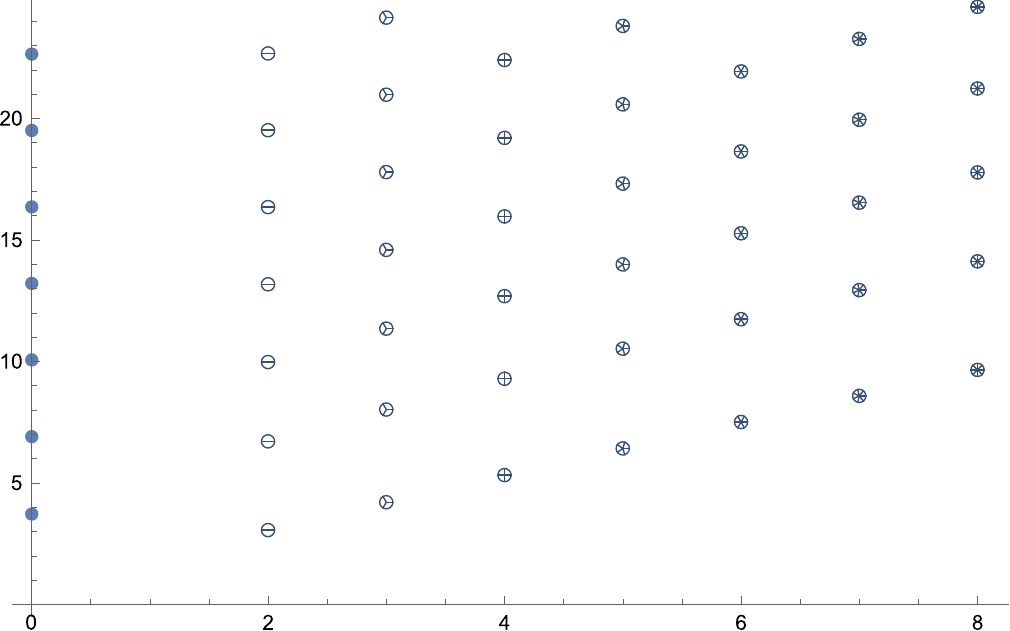}}
 \vspace{-1mm}
 \caption{Sloshing eigenvalues for circular, infinitely deep cylinders with $m$ radial,
  porous (rigid) baffles: any number of arbitrarily distributed baffles when $n=0$; a
  diametral baffle when $m n = 2$; then $3,\dots,8$ baffles with equal angles between
  them for the rest values of $m n$.}
 \label{fig:fig3}
\end{figure}

Similar to \eqref{u'}, explicit solutions analogous to \eqref{u4} and \eqref{com4}
have the form
\begin{equation}
\phi_{n,s}^{(m)} (r, \theta, z) = u_{n,s}^{(m)} (r, \theta) \, \E^{k_{m n,s} z} \ \
\mbox{and} \ \ \nu_{n,s}^{(m)} = k_{m n,s} , \ \ n = 0,1,2,\dots, \ s = 1,2,\dots,
\label{u4i}
\end{equation}
in the case of infinite depth. The latter eigenvalues are illustrated in fig.~5. The
values $\nu_{0,s}^{(m)}$, $s = 1,2,\dots$, given in the leftmost column describe
sloshing frequencies of purely radial modes in the infinitely deep cylinder of the
unit radius with any number $m \geq 2$ of porous (rigid) baffles connecting the axis
and the circular wall, but distributed arbitrarily on $[0, 2 \pi)$. In the next
column, the values $\nu_{1,s}^{(2)}$, $s = 1,\dots,7$, are given; they describe the
sloshing frequencies in the cylinder with a diametral porous (rigid) baffle. Each of
the rest six columns contain the values $\nu_{n,s}^{(m)}$ for some value of $s$ in
every column and $m n = 3,\dots,8$; they describe the sloshing frequencies in the
cylinder with radial porous (rigid) baffles with equal angles between them. Of
course, one has the equality $\nu_{n,s}^{(m)} = \nu_{m,s}^{(n)}$ for this kind of
eigenvalues. It is easy to find several pairs located at the same level in fig.~5.

It is easy to obtain formulae analogous to \eqref{u4}--\eqref{u4i} for annular
cylinders with baffles dividing them into $m$ parts, but instead of $k_{m n,s}$ the
corresponding values must be determined in the same way as in sect.~4.

\section{Concluding remarks}

In this note, examples of explicit solutions to the sloshing problem in containers
with porous baffles have been constructed; to the best authors' knowledge no such
examples were known so far. Vertical cylinders with circular walls occupied by an
inviscid, incompressible, heavy fluid and containing radial baffles stretched
throughout the depth were considered and the fluid transmission across the baffles
was described by the widely used Darcy's law.

The crucial feature of the obtained explicit solutions is that the corresponding
pressure difference vanishes across the baffle itself for each of them. Therefore,
these solutions coincide with those that describe sloshing in the case of rigid
baffles which means that in these situations porous baffles with any $\beta$
characterizing the porosity of the baffle are equivalent to the rigid ones having
the same configuration. The damping efficiency of the latter baffles is due to
changes of the velocity field in the fluid and this was investigated by the authors
earlier; see, for example, \cite[fig.~6]{KM}. Hence the damping efficiency of porous
baffles is the same as that of rigid ones at the frequencies used in the considered
examples. Finally, the obtained eigenfrequencies are real, and so the corresponding
sloshing oscillations are non-decaying in time. At the same time, the problem's
formulation is complex valued and this suggests that generally speaking oscillations
should decay exponentially in time; see, for example, the model investigated in
\cite{Tu}.

An important point concerning further investigation of these examples is to find out
whether the corresponding eigenfrequencies are simple for porous baffles as it takes
place for rigid ones. Future work of potential interest is to consider whether a
characteristic equation can be derived for the unknown frequency $\omega$ in
problem, say, \eqref{slosh1}--\eqref{slosh5}; then it can be solved numerically as
in the paper \cite{Tu}, where a container of different geometry was considered.
Finally, it would be interesting to study whether there are explicit sloshing
solutions for containers with porous baffles other than those considered here.

In the light of obtained explicit solutions, it is reasonable to question whether
Darcy's law (see its derivation and extensions in \cite[ch.~1]{book}) provides a
suitable boundary condition to be posed on a porous baffle in the sloshing problem.
Numerous limitations on applicability of Darcy's law and even its breakdown were
discussed in the literature (see, for example, \cite{rev} and \cite{soft}); in
particular, it was noted that its use in the case of preferential flow pathways is
problematic. However, such a pathway indeed happens in the case of a baffle;
therefore, no wander that paradoxical results have been obtained for various
containers. Namely, oscillation modes are independent of properties of the porous
medium and coincide with those for rigid baffles. The question how to modify
Darcy's law in the baffle boundary condition is of paramount importance for
obtaining plausible results in this important problem.

\end{document}